\documentclass[10pt,conference]{IEEEtran}

\usepackage{cite}
\ifCLASSINFOpdf
 \usepackage[pdftex]{graphicx}
\usepackage{amssymb}
\usepackage{lineno}
\usepackage{lscape}
\usepackage{tabularx}
\usepackage[gen]{eurosym}
\usepackage{multirow}
\usepackage{hhline}
\usepackage{booktabs}
\usepackage{siunitx}
\usepackage{float}
\usepackage{framed}
\usepackage{url}
\usepackage{amsmath}
\else
\fi

\begin{document}
%
\title{Social Identity in Software Development}

\author{\IEEEauthorblockN{Andreas B\"{a}ckevik}
\IEEEauthorblockA{\emph{SKIM Group}, Gothenburg, Sweden, and\\ \emph{Chalmers $|$ University of Gothenburg}\\ Gothenburg, Sweden\\
Email: a.backevik@skimgroup.com}
\and
\IEEEauthorblockN{Erik Thol\'{e}n}
\IEEEauthorblockA{\emph{SKIM Group}, Gothenburg, Sweden, and\\ \emph{Chalmers $|$ University of Gothenburg} \\ Gothenburg, Sweden\\
Email: e.tholen@skimgroup.com}
\and
\IEEEauthorblockN{Lucas Gren}
\IEEEauthorblockA{\emph{Chalmers $|$ University of Gothenburg}\\ Gothenburg, Sweden\\
Email: lucas.gren@cse.gu.se}

}
\maketitle

\begin{abstract}
An agile approach has become very popular over the last decade, which requires good communication and teamwork within teams as well as with outside stakeholders. Therefore, social interaction is central for a software development team to be successful. Such social interactions form social identities and social structures in both teams and organizations. This study investigates possible effects that the social identity of individuals may have on the effectiveness of software development through seven in-dept interviews. The qualitative data from interviews were analyzed and summarized using summative content analysis, and the seven individuals also answered a questionnaire on social identity taken from social psychology research. The qualitative result shows that aspects of social identity affect software developers' behavior, and that we need to build cross-functional stable teams over time also from a pure social identity perspective in addition to the product related aspects to avoid a decreased effectiveness. However, we did not see clear connections to our operationalization of effectiveness in this study, and the quantitative analysis was also inconclusive, but we see value in our suggested method when investigating social identity in software development.

\end{abstract}

\IEEEpeerreviewmaketitle

\section{Introduction}
In many fields, projects are commonly carried out in teams, mainly due to the belief that teams empower individuals to be more productive \cite{positive_workplace}. Most agile approaches have originated from the agile manifesto \cite{agile_manifesto}, which states ``Individuals and interactions over processes and tools'' that stresses the importance of interaction within and between teams.

In an agile team, group maturity affects different aspects of team agility \cite{group_development_gren}. According to Wheelan \cite{group_development_wheelan_time}, groups need on average six and a half months to become high performing. Gren \cite{psychological_group_processes} suggests applying social identity theory to find more complex correlations between team maturity and team agility in agile projects.

Social identity theory describes how individuals relate to and reflect on social groups they belong to. For example, Abrams and Hogg \cite{social_identity_theory} describe an extreme where interactions between multiple individuals can be solely based on social groups they belong to and not interpersonal relations. To be able to describe and quantify an individual's social identity within groups, Luthanen and Crocker \cite{collective_self_esteem_scale} created a scale to measure collective self-esteem which is divided into four aspects. By using this scale while also measuring team performance, potential relationships between one's social identity and team performance can be explored.

A group's performance can be defined very differently based on the context of its surroundings. For instance, a customer service company may define a team's performance based on customer ratings and not speed. Cambridge Dictionary's \cite{effectiveness_def} definition of effectiveness is ``the ability to be successful and produce the intended results.''

The objective of this paper is to investigate the potential impact of social identity on effectiveness in teams in the field of software engineering. Semi-structured interviews were conducted to explain individuals' social identity and team effectiveness. As an addition to the qualitative data, social identity was measured through a Collective Self Esteem scale \cite{collective_self_esteem_scale}. The questionnaire was extended to ask for the respondents' effectiveness to bundle it with their social identity measure. This is one of the first studies that connects social identity theory to software engineering, but we study social identity based on the assumption that it does play a role of some magnitude in software engineering, since people write the software. The research question is therefore: \emph{How does the social identity of software engineers relate to software development effectiveness?}


\paragraph{Personal Identity}
According to Stets and Carter \cite{personal_identity_theory}, the core of an identity is how one categorizes oneself into a specific role and how it is incorporated by the person. This includes meanings and expectations within that role, such as performance and behavior. For example, individuals who work in law enforcement, will likely categorize themselves into that specific role and expect certain behavior from themselves such as being helpful and just. An individual's expectations on a specific role form a standard that guides further behavior within and around all roles. Stets and Burke \cite{identity_theory} explain interaction as an important component in roles, and much of the activity within a role from individuals revolve around these interactions. All of these interactions within and outside roles define our social structure. In short, an individual who strongly identifies with a specific role will try to fulfill expectations, manage interactions and control the environment that the role is responsible for. 

Besides categorizing oneself into a specific role, negotiated roles are also evident in identity theory. A team captain in a sports team is an example of a negotiated role. Research has found that individuals with these types of roles become less satisfied when their social group could not verify their identity \cite{identity_theory}. As groups affect an individual's role and how they categorize themselves, there is a correlation between personal identity theory and social identity theory \cite{identity_theory}. 

Reasons for why individuals commit to activating a specific identity have been widely discussed by researchers. Stryker and Serpe \cite{identity_salience} split it up into two factors: the number of individuals that a person is tied to through an identity and strength in ties to others through an identity. The more an identity is embedded into individuals in a social structure, the more likely it is that the identity will be activated in specific situations. On the other hand, the stronger ties an individual has through an identity the more it leads to a more salient identity, which can become a characteristic of the identity and not be based on the situation. In this way, one can differentiate the probability of an identity being used in a situation or if a salient identity is activated \cite{identity_theory}. 

\paragraph{Social Identity}
While personal identity theory focuses on specific roles and identities' interaction in a social group, social identity instead focuses on social roles within social categories and social groups \cite{social_identity_theory}. Hogg et al. \cite{social_identity_perspective} describe a social identity as ``a person's knowledge that he or she belongs to a social category or group.'' Luthanen and Crocker \cite{collective_self_esteem_scale} exemplify this with the assessment ``how attractive one feels,'' positioning oneself as a member of the group ``attractive people,'' or ``non-attractive people.'' Hogg et al. \cite{social_identity_perspective} also described a social group as ``a set of individuals who hold a common social identification or view themselves as members of the same social category.'' People inside a specific social group are often referred to as in-groups, while people outside such groups are referred to as out-groups \cite{hogg2014sp}.

During formation of social identity, two important processes are involved. These are social comparison and self-categorization and they have different consequences on one's social identity \cite{social_formation}. During self-categorization, an individual emphasizes perceived similarities with the in-group while also emphasizing perceived differences with the out-group. The accentuation regards to attitude, beliefs, behavior and other similar properties \cite{social_formation}.

Abrams and Hogg \cite{social_identity_theory} also argue that an individual strives to have a positive social identity by comparing with others in their social group or outside. Evaluation of social groups is done by comparing them to other social groups with similar characteristics. However, when an individual is not satisfied with their social identity, they tend to leave their existing social groups to find more positive ones. Therefore, not being able to leave a social group that gives one a negative social identity, will likely lower ones collective self-esteem \cite{social_identity_theory,social_identity_perspective}. Luthanen and Crocker \cite{collective_self_esteem_scale} created a thoroughly validated scale that describes and assesses parts of an individual's social identity, which focuses on the collective self-esteem part. It is categorized into four sub-scales: membership esteem, public collective self-esteem, private collective self-esteem and importance to identity. Membership esteem refers to how an individual sees oneself in a group. This aspect assesses the most individualistic aspects of one's collective self-esteem (CSelfE) \cite{collective_self_esteem_scale}. Public collective self-esteem refers to how the individual's group is evaluated by others, which assesses one's judgment of how other people evaluate the group. Private collective self-esteem refers to how an individual evaluates the group one belongs to, which assesses one's personal judgment of how good the group is. Importance to identity refers to how important group membership is to an individual.

\paragraph{Group Effectiveness}
It is difficult to find an operationalization of effectiveness in the software engineering domain, but Al-Sabbagh and Gren \cite{khaled} argue that the degree of how well a team can plan for future iterations (or sprints) is a measurement of effectiveness in the software engineering context, since it describe if the team could deliverer the intended and planned result to the customer. Therefore, this paper looks closer into the planning effectiveness of social groups within software development due to effectiveness being more tangible than group performance. We measure team effectiveness in this study as estimated effort divided by spent effort, just like in Al-Sabbagh and Gren \cite{khaled}.

The teams under investigation in this study break down stakeholder requirements into tangible tasks along with an effort estimate in hours. Eventually, these tasks will be completed and given an actual effort in hours that it took to complete it, and the effectiveness can be seen as the estimated effort divided by actual effort.

\section{Method}

\subsection{Participants}
The software development teams within the organization under investigation were split into two subgroups: support and development. The purpose of a support group was for one software developer to aid with technical support to a non-developer project team. Support groups were often short-lived (a couple of hours up to one week) and could include technical tasks such as debugging or developing small applications with a very specific purpose. Development groups consisted of only software developers. The goal of development groups was to develop and deliver products. Several projects could be ongoing in parallel, which meant that software developers could be a part of several development groups at the same time. There was no consideration of cultural differences based on different geographical locations, nor to any personal factors such as gender, race, religion, nationality, ethnicity and socioeconomic class, to keeping the participants' anonymity.

All participants were software developers at the same company. However, one participant was a lead developer and had management responsibilities. Four of the participants had worked less than a year at the company, while the other three had worked at the company for two to four years.


\def \nbrOfInterviews {7}

\subsection{Qualitative Data}
The main source of empirical data of this paper is semi-structured interviews. A total of \nbrOfInterviews{} in-depth interviews were conducted. An interview guide was constructed with the CSelfE scale \cite{collective_self_esteem_scale} as a basis. The interview guide aimed at obtaining information on how the interview participants related to the four aspects of collective self-esteem (\emph{Private collective self-esteem}, \emph{Public collective self-esteem}, \emph{Membership esteem}, \emph{Importance}).


We wanted the interview to be flexible and to find any possible explanation of the respondent's social identity. Thus, we chose to have the interview as a semi-structured interview that further enables emerging of new concepts \cite{interview_preparation_conduction}. 

The interview guide was constructed and reviewed in iterations. A total of three test interviews were held with participants that did not have any relation with the organization under investigation for the study. For this reason, the specific teams and situations that referred to the organization were replaced with something more relevant for the participants. After reviewing the last test interview, the data was similar enough to the previous one that we felt ready to use the guide in a real scenario.

With the main goal of the interview being to understand the social identity of the respondent, questions were constructed as open questions with narrowing probes prepared. For example, we predicted that respondents would relate differently to the support groups than how they related to their usual development group collaboration instances. To collect data to support this prediction, we asked respondents about challenges that they faced in their daily work and if they did not mention the support groups, we would ask them if they saw those as a challenge. We used similar probing questions in many places to answer questions related to the social identity, such as how the respondents thought that others within the organization viewed their teams.

The introduction stated the purpose of the interview and expressed how the moderator has no personal interest in the discussions that would take place. We also clarified the terminology to make sure that the respondent considered effectiveness and teams as closely as possible to how we defined it for the study. Lastly, the respondent was ensured that their identity would be confidential and asked for permission to record the interview.




\subsection{Quantitative Data}

While the current study is mainly focused on qualitative data, we considered quantitative data to be of value as it would be able to support our findings and conclusions drawn from the qualitative data. The quantitative data also helped in developing a common vocabulary and communicate our goals to the participants in the study. 

\paragraph{Self-Esteem Questionnaire}
As stated in the introduction, the scale used to measure collective self-esteem was constructed by Luthanan and Crocker \cite{collective_self_esteem_scale} and assesses four aspects of collective self-esteem: (1) Private collective self-esteem, (2) Membership esteem, (3) Public collective self-esteem, and (4) Importance to identity. Each aspect is assessed by four items, resulting in a total of 16 items in the scale. The responses to these were made on a 7-point Likert-type scale.

For the purpose of this study, we modified the original instructions to instruct the respondent to consider different groups that better represent the software development teams present within the organization under investigation. We constructed two versions of these instructions, one for support and one for non-support projects. The creators of the scale, Luther and Crocke \cite{collective_self_esteem_scale}, have validated the scale and questions that belong to the CSelfE questionnaire. Therefore, we did not have to validate each question, but only to adapt the context of each question to fit the narrative of this study.

\paragraph{Team Effectiveness}

We calculated team effectiveness as equal to earned over planned effort, which refers to the estimated effort that would be spent and earned refers to the amount of effort spent during a collaboration instance. These values were readily available due to the nature of how they planned and managed their software development tasks. 

We chose not to include any preconceived unit for these values, as they could represent whatever the team deemed fitting for their way of working. For non-support collaborations, these are likely to be story points or other subjective estimates \cite{effort_estimation} while for support collaborations we would expect actual hours due to the ad hoc nature of these collaborations within the organization. As we are interested in the ratio of the estimated and spent effort we disregarded the unit and any implications it would have on the project. Thus, we allowed the participants to interpret the question freely and put whatever unit if any, that they preferred.

\paragraph{Data Collection}
An online survey was constructed, and links to this survey were then sent out to the subjects. The survey consisted of a series of questions where we would receive the following data for each response: (1) Collaboration instance type (support team or software development team), (2) CSelfE questionnaire, (3) Planned effort for the collaboration instance, and (4) Spent effort for the collaboration instance.

A total of seven participants took part in the study, which could of course be seen as very few. However, conducting in-depth interviews regarding the social identity of software developers is a time-consuming endeavor. Since we also wanted to use the questionnaire developed by Luthanen and Crocker \cite{collective_self_esteem_scale}, the quantitative result should only be seen as a suggestion of how to investigate these connections on a larger scale. 

At the end of each collaboration instance, we asked the participants to take the survey. Weekly reminders were also sent out in an attempt to prevent omitted responses. Individuals were reminded daily regarding the support team questionnaire, while they were reminded weekly regarding the software development team questionnaire.

\section{Data Analysis}
The interview recordings were transcribed separately and then merged into one for each interview. The reason for this was to even out potential differences in how we would interpret the recordings. According to Hsieh and Shannon \cite{approaches_qualitative_content_analysis}, keywords in a summative content analysis are derived from interests of the researchers while other approaches to content analysis are derived from the data or theory. As our interest was to understand the social identity of respondents, we used a summative approach to qualitative content analysis.

We analyzed the final transcripts together (the first and second author), looking for anything that could be tied to any of the four aspects of CSelfE in addition to planning effectiveness. Any statement that was deemed to belong to any of the aspects was then labeled with a keyword in consensus where we would discuss what keyword that statement expressed or represented. This yielded an aggregate of grouped keywords that appeared several times within one or many aspects which could be handled as quantitative data.

The collective self-esteem scale developed by Luthanan and Crocker \cite{collective_self_esteem_scale} provides four questions per identity category. In accordance to previous research using that tool we took an average of these four questions as a measure of a social identity category. Some of the questions should be reversely scored, which we also followed in our calculations.

Quantitative data was gathered from the seven respondents, who also took part in the interviews, where each respondent was asked to participate in the survey at least once a week. As we collected data for one month, this would yield 28 responses at best. With a response rate of \begin{math}53\%\end{math} a total of 15 survey responses were gathered. 

We would like to highlight here that only 15 responses to all these questions regarding social identity is very small. We will show the data analysis we conducted below mostly to suggest how that method could be applied to larger samples in the future. However, since the results were statistically non-significant this could also mean that there are not any connections between social identity and planing effectiveness, which then contradicts the qualitative results. We believe, though, that more data is needed for such a conclusion.

Since we advocate using Bayesian data analysis (BDA) for all statistics, we will use such tools as provided and implemented by McElreath \cite{McElreath2016sra} using R \cite{R} and Stan \cite{carpenter2017stan}. There are several advantages of using BDA and there is a myriad of textbooks and scientific articles arguing for the use of such an approach (e.g.\ \cite{van2014gentle}). Briefly, the advantages include skipping non-intuitive null hypothesis testing, using the real distributions of data in connection to explicitly stating the distributive assumptions, adding distributions to all parameters instead of using point estimates, using prior information about parameters when new data is collected, etc.\ \cite{McElreath2016sra}.

The likelihood functions and our weakly informative priors \cite{bernardo1975non} used when the first data was analyzed were the following:
{\footnotesize
\begin{equation}
\mathrm{Effectiveness}_i  \sim  \mathrm{Normal}(\mu_i, \sigma)
\end{equation}
\begin{equation}
 \begin{aligned}
    \mathrm \mu_i  =  \alpha + \beta_{M} \mathrm{Membership}_i + \beta_{Pr} \mathrm{Private}_i +\\
      + \beta_{Pu} \mathrm{Public}_i + \beta_{I} \mathrm{Importance}_i
  \end{aligned}
\end{equation}
\begin{equation}
\alpha  \sim  \mathrm{Normal}(0,5) 
\end{equation}
\begin{equation}
\beta_{M}  \sim  \mathrm{Normal}(3.5,3)
\end{equation}
\begin{equation}
\beta_{Pr}  \sim  \mathrm{Normal}(3.5,3)
\end{equation}
\begin{equation}
\beta_{Pu}  \sim  \mathrm{Normal}(3.5,3)
\end{equation}
\begin{equation}
\beta_{I}  \sim  \mathrm{Normal}(3.5,3)
\end{equation}
\begin{equation}
\sigma  \sim  \mathrm{Cauchy}(0,2)
\end{equation}
}
The priors above need some explanation. The response variable is always assumed to be Gaussian (i.e.\ normally distributed) in linear regression \cite{McElreath2016sra} which is why our estimate variable is assumed to be Gaussian with a $\mu_i$, and $\sigma$. The first four rows are our statistical model saying that the mean values are equal to an intercept and our research variables for each social identity category. When using and BDA and explicitly defining our statistical model like this, makes it possible to directly observe our hypothesis about the data since we use our subjective knowledge as priors in the statistical model. In our case, we do not know much about the prior distribution, however, we think that the social identity factors should be larger than zero and not vary more than a standard deviation of 3 on a Likert scale of 1--7. This value is quite arbitrary but we model our uncertainty about this value by setting the standard deviation of the variables to 3. The value of the means for the prior distributions have very little impact on the result, which was also confirmed when running the analysis with other mean values.

The $\sigma$, i.e.\ the standard deviation of the distributions, were assumed to be quite heavy tailed, and therefore considered as a weakly informative, and a standard prior for standard deviations \cite{gelman2008weakly}.

\def \nbrOfInterviews {seven}
\section{Results}\label{results_section}
The interview was constructed with each aspect of CSelfE in mind. Below follows a summary of the responses for the \emph{Private collective self-esteem}, \emph{Public collective self-esteem}, \emph{Membership esteem} and \emph{Importance to identity} aspect.

\paragraph{Private collective self-esteem}
In general, the seven respondents evaluated their established group of fellow software developers positively, as 75\% of the identified keywords were positive. The four keywords: expertise, team spirit and communication made up for 30\%, 28\% and 24\% of the identified positive keywords respectively (see Figure 2\footnote{Raw data available at: \url{http://bit.ly/2IBWAPd}}). When asked to describe how their teams progressed towards their goals, one respondent positively expressed the following about a development team collaboration instance which summarizes the respondents' opinions:

\begin{quote}
``We talk things through and estimate together to visualize a common goal. We combine different competencies and experiences to find the best way of doing things, which is due to prestigeless communication. You can speak freely.''
\end{quote}

All seven respondents considered the different personalities within their development group collaboration instances to work well together in and that this enabled good communication within the team. The respondents would also state that the environment within the team was relaxed.

During the assessment of their support teams, communication was mentioned by four respondents as a negative factor. A lack of understanding and differences in technical competence often caused agitation and stress. When asked about the biggest challenge of working as a software developer within the company, six out of \nbrOfInterviews{} respondents mentioned meeting deadlines, planning and communicating with managers. Difficulties in communication with individuals outside of their teams or with less experience in software development was also a recurring topic. 

\begin{figure*}\label{keyword_distribution}
  \centering
    \includegraphics[width=5.4in]{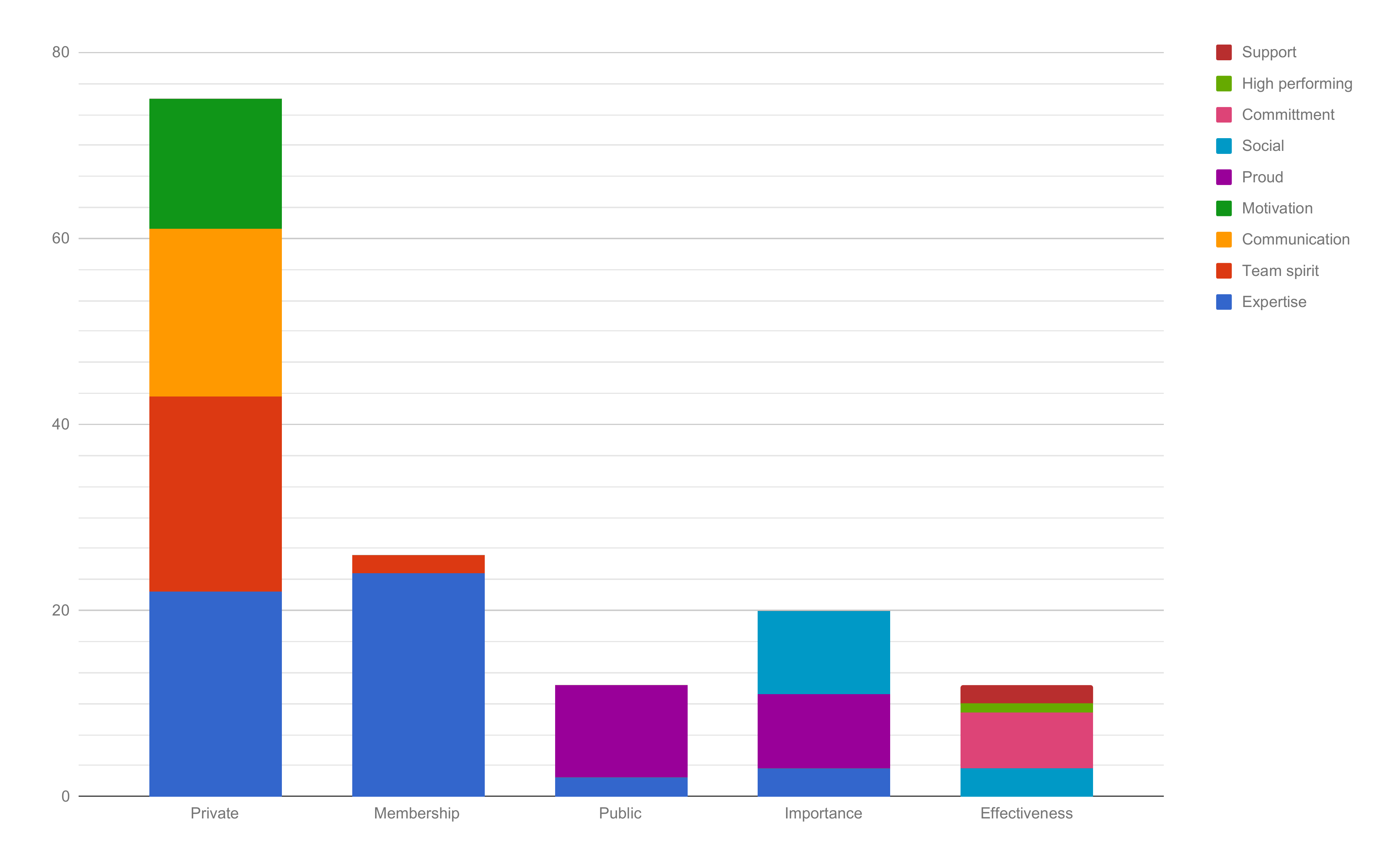}
    \caption{Positive Keyword distribution}
\end{figure*}

\paragraph{Public collective self-esteem}
When assessing how the respondents think others evaluate their teams, all seven respondents felt proud of their combined expertise and thought that their competence was recognized by others. All twelve identified positive keywords found regarding public evaluation were either \emph{pride} or \emph{expertise}. However, a total of 24 negative keywords were grouped under \emph{Public collective self-esteem}, rendering a majority of the keywords found for \emph{Public collective self-esteem} as negative. The respondents thought that others viewed them as arrogant, unproductive or uncommitted. Most respondents mention lack of commitment, friction and lack of understanding as a result of the negative view of their teams. One respondent said the following:

\begin{quote}
``As developers, we are guests in the environment of the analysts.''  
\end{quote}

This statement summarizes what was described by the respondents on how it is to be a software developer within the organization. There is a clear gap between software developers and non-software developers, where communication and understanding are lacking in collaboration instances where non-software developers take part. Even so, previous accomplishments made by the software developer teams were assumed by respondents to contribute to a more positive public image.

\paragraph{Membership esteem}
Throughout the interviews, all \nbrOfInterviews{} respondents mentioned that they contributed with technical expertise to all of their collaboration instances they were a part of. However, no respondent mentioned their own technical expertise as a factor to team effectiveness and only two respondents mentioned their own expertise more than once. In general, the respondents were content with their team membership but did not focus on it in any question.

As the respondents had a clear view of their \emph{Membership esteem} but did not express any specific feelings related to it, we could not find any association to either effectiveness or any other keywords mentioned throughout the interviews.

\paragraph{Importance to identity}
The two positive keywords identified were \emph{socializing} and \emph{pride} (see Figure 1) both occurring equally many times. Four out of seven respondents considered the social aspect of group belonging important. The respondents spoke about their development groups when discussing the importance and not their support groups. The importance of one's self-concept of belonging to teams was stronger for the established teams and weaker when dealing with support group collaboration instances.

All \nbrOfInterviews{} respondents were proud of their software development teams. The culture of these development teams and their accomplishments were a few things that made respondents feel pride towards them. It was also brought up that if others, outside of the teams, would recognize their accomplishments they would feel even more proud.

\paragraph{Software Development Effectiveness}
A few areas of improvement were brought to light from the interviews. The requirements for projects are often too few, too vague and they should be more detailed and clear to everyone involved in the project. The respondents thought that this could be resolved by additional communication with stakeholders in order to resolve any uncertainties.

In general, when evaluating effectiveness, respondents differentiated support teams from their development teams. They believed that projects executed within their development teams were less effective than the support teams, mainly due to the fact that projects taken on by development teams require more complex engagement and communication with stakeholders. Stakeholders were considered to be reactive rather than proactive which led to the end goals changing. The development teams would try to work around this by communicating with the team and sort out any internal uncertainties and, in the end, work towards the same goal with best possible effort.

Five respondents considered themselves to excel in the problem-solving aspects of working in short-term support groups. The problems to be solved in these projects were however often not stimulating to the respondents. When the respondents were asked if they considered their teams to be high performing, they instead focused on development teams. Six out of \nbrOfInterviews{} assessed that they are performing average or slightly above average. The respondents also mentioned that work in support groups was very effective. This was due to work in support teams being transactional and easy to plan. Thus, we conclude that development teams are perceived as less effective but that they do solve much more complex tasks.


\subsection{Analysis of Quantitative Data}
\begin{figure}
\includegraphics[scale=0.25]{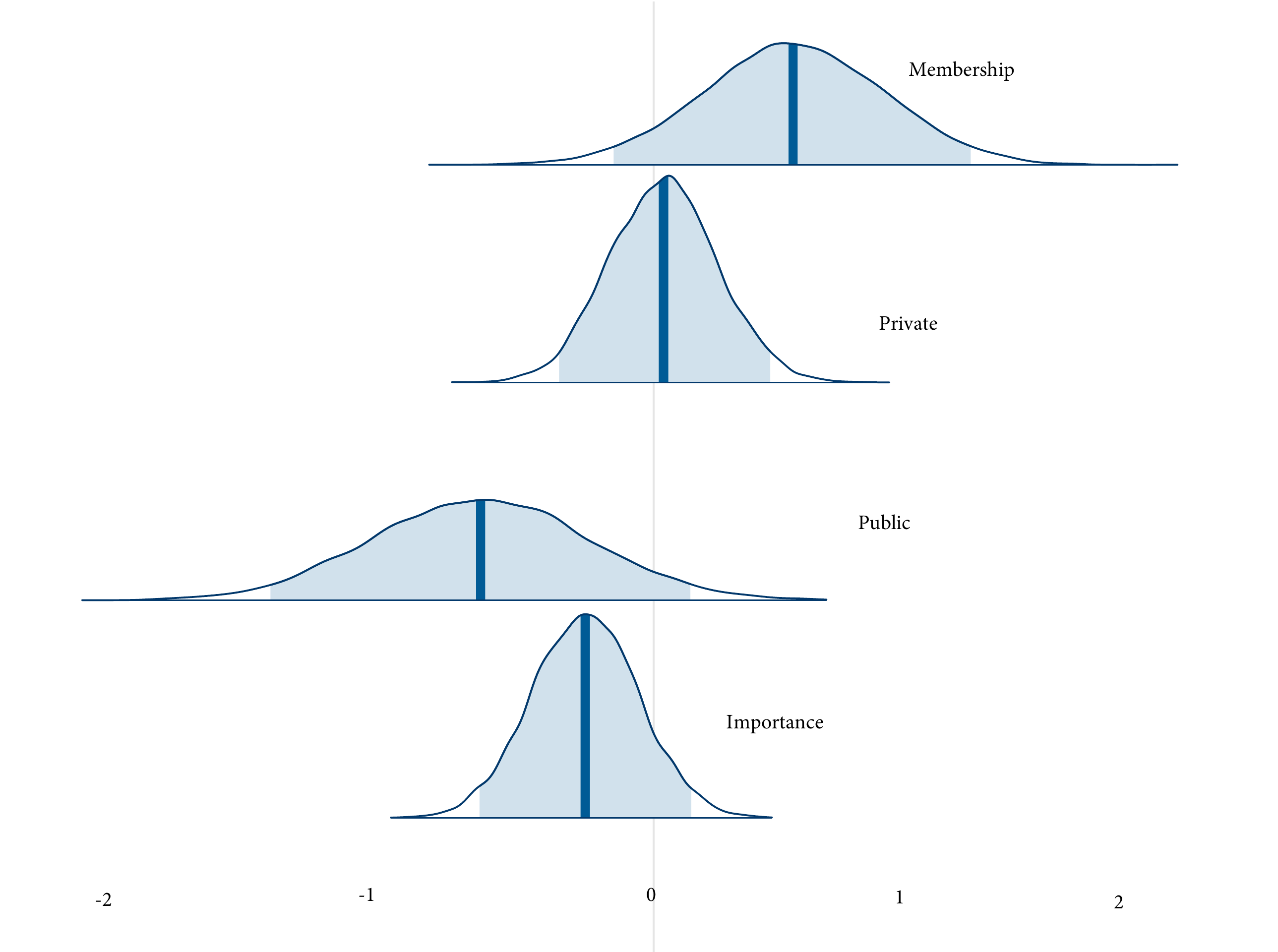}
\caption{Sampled posterior distributions (based on 10k samples) for social identity factors with median and 95\% credible interval.}
\label{fig:1} 
\end{figure}

Figure \ref{fig:1} shows the sampled posterior distributions with a connected 95\% credible interval, which shows that none of the factors are significantly different from zero.

The posterior distribution is calculated from combining the likelihood functions with our collected data. It is rarely possible to exactly calculate the posterior distribution, which is why we instead sampled 10k samples from the posterior that were then used in the analysis. This is one of the reasons BDA was less used before modern computers with enough computing power for such sampling methods \cite{McElreath2016sra}. Using BDA in software engineering is quite novel, and for more details in the use of BDA in software engineering, see \cite{2018arXiv180909849T}.

By using our BDA, the definitions of the statistical models and the way we obtain the posterior distributions are more complex, however, now that we have these distributions the conclusions drawn in relation to our research is extremely straightforward and require no knowledge of statistical assumptions, non-intuitive interpretation of $p$ values or failing to reject the hypothesis we did not believe initially (the reader probably gets the point made here). By simply looking at Figure~\ref{fig:1}, we see that no factors were significantly different from zero, i.e.\ they do not significantly explain variance in planning effectiveness.

\section{Discussion}



\subsection{Private collective self-esteem}
The software developers within the organization chose to focus on their most established teams (development teams) which consisted of only software developers. They evaluated their development teams positively, motivated by the fact that these groups are often made up by other software developers. However, such collaboration instances often contain interactions between software developers and stakeholders. As stated earlier, a group is a set of people who work towards a common goal which would include both developers and stakeholders in one team. No respondent seemed to identify stakeholders as members of their social groups when evaluating their development groups. There is a clear social structure that posits developers and non-developers as distinctly different social groups within the company, which strengthens the respondents social categorization. The software developers' categorization of themselves, non-developers and stakeholders likely has an effect on the end result as communicative issues might arise between the groups. We were unable to find any connection between effectiveness and the private aspect of CSelfE in either the qualitative data or the quantitative data. 

It is, however, evident that the software developers scored higher on the \emph{Private collective self-esteem} aspect and had a more positive view of their development teams than their support teams. Based on the interviews, we believe that the reason for this is the fact that software developers are able to strengthen the positive image of their group when surrounded by others like themselves. Their expertise is recognized and valued higher as understanding is higher among fellow software developers than among less technical individuals, which was the case in support groups. The fact that the support groups were not as positively evaluated as the development teams could be a result of a favorable comparison where software developers try to strengthen the positive image that they have of their development teams. It is possible that this comparison is done on a level where the software developers justify their social group ``software developers'' and not the group within the organization -- the development team -- as they effectively compare software developers to non-software developers.

If we want developers to be part of cross-functional agile teams, having them move around between teams seems to distance them from other roles, which, since agile teams overlap with high performing teams \cite{group_development_gren}, is a great disadvantage. The present study highlight the importance of cross-functionality in teams over time in relation to social identity, which adds to the knowledge of why it is such an advantage.

\subsection{Public collective self-esteem}
Each software developer considered all other software developers in their teams to have high technical expertise, which enabled them to feel proud that they were viewed as skilled by out-groups. However, Section \ref{results_section} also concludes that there is agitation towards the software developers, in which non-developers occasionally thought of software developers as arrogant. 

According to the software developers, the negative image is created by miscommunication and lack of understanding which could be the result of the strong social categorization mentioned earlier. One software developer mentioned that task estimates are done in precaution to safeguard from over-committing. This is likely an effect of bad communication, previous misunderstandings and negatively impacts projects which are often planned with respect to estimates. If these estimates are not true or intentionally set higher, overall effectiveness suffers. It is possible that this effect is present in our data, which would render how \emph{Effectiveness} is measured in this paper as a poor reflection of the true effectiveness. A possible cause of this effect could be the preservation of a positive social identity, a measure not far from the predictions made by Luthanan and Crocker \cite{collective_self_esteem_scale} where high collective self-esteem individuals were predicted to engage in competition with out-groups in the face of collective threats.

From this, we conclude that a connection between \emph{Public collective self-esteem} and \emph{Effectiveness} may be present in that possible actions taken to counter a negative \emph{Public collective self-esteem} assessment could influence estimates and possibly other elements of software development. Estimation is related to planning and in our paper, the planned effectiveness does not fully capture effects on all elements of software development, such as requirements and communication. Therefore, one should look further into measurements of \emph{Effectiveness} to confirm or reject a potential correlation with \emph{Public collective self-esteem}.

\subsection{Importance to identity}\label{conclusion_importance}
No software developers considered temporary support teams as important to themselves, probably because they included other roles but, of course, also because they were temporary. One software developer mentioned that the temporary support groups were more effective because it was very transactional and required little communication and interaction compared to the established teams. We believe that the software developers distance themselves from non-developer groups, due to previous misunderstandings and miscommunication, to be able to focus on problem-solving. Therefore, software developers assessed support teams as more effective than their development teams. Thus, we believe that \emph{Importance to identity} is connected to \emph{Effectiveness}, but with communication as an important determinant. However, we were unable find any correlation between \emph{Importance to identity} and planned \emph{Effectiveness} in the quantitative data, possibly due to the small sample size. We conclude, though, that it seems to be a better idea to leverage cross-functionality in teams over time also from a social identity perspective. 

\subsection{Answer to Research Question}
\emph{How does the social identity of software engineers relate to software development effectiveness?}
We are confident that there is some interplay between \emph{Private collective self-esteem} and \emph{Public collective self-esteem} in addition to \emph{Importance to one's personal identity} based on the qualitative results, and that this interplay must somehow affect the effectiveness of the work in relation to deliver value to their customers. However, using effectiveness in the way we used it this study did not capture the connections between these constructs explicitly. Individuals seem to protect and defend their groups from a negative public view to differing extents, affected by a combination of their private evaluation of the groups and how important they are to them. If their \emph{Private collective self-esteem} is low within a certain group, it is likely that they do not consider it to be of high \emph{Importance to their identity} and the magnitude of these measures is lower as they are able to handle a negative public view without as much impact on their social identity. If it is not important to them, they do not value what others think about it. Unfortunately, this has an effect on the quality of communication which in turn feeds back into the cycle and the software development is less effective as a result of this. With developers being distant to their team-members if they are not all developers must have an effect on the teams' ability to deliver value to the customers, given that the values is probably a bit more complex than basic technical functionality with clear requirements.

In social psychology, Dovidio et al. \cite{dovidio1998intergroup} have shown that it does not help to tone down the differences in identities, but it is way more effective to include people from different groups on the basis that the roles are defined as different and maintained with their positive distinctiveness when cooperating. The developers and the other employees then need to define projects together based on what is needed for goal-fulfillment. If the team then together realizes that all the different competences are needed in the team, they could celebrate their different skills but still work toward a common goal as one in-group. In this way our present study supports the use of DevOps in that maintenance tasks and development tasks need to be combined in the same team, but also the importance of cross-functionality in teams also from a pure social identity perspective in order to increase the software development effectiveness.

\subsection{Threats to Validity}
Talking about one's collective self-esteem and social identity can be sensitive and touch upon matters that one is not comfortable talking about. Therefore, as two of the authors of this paper were working in the same organization as the respondents, their answers may be biased or incomplete. As the interviewees could not be anonymous, we explained that we were conducting research as scientists, and not fellow employees. However, there is no way to reassure that every respondent answered truthfully, which could threaten the validity of our data.

As no good measures were available to measure effectiveness, we used effort estimation to quantitatively measure effectiveness. Even if we believe planning effectiveness captures more than just estimation accuracy of features, it does not always relate to customer value. In our case, the effort estimation was probably more a predictor of \emph{Efficiency} (doing something well) rather than \emph{Effectiveness} (doing the right thing). Our results were also clouded by the fact that the support groups had much less complex projects with requirements that were easy to estimate and break down. 

This study should also be replicated with a much larger and more diverse sample in order to verify our findings and conclusions. In a more diverse sample, with software developers across multiple organizations, company culture effects can possibly be identified. As this paper only surveyed software developers within one organization, it could be that such aspects affected the results. Furthermore, a larger sample would allow for a more robust quantitative analysis where we, on the contrary, had a total of 15 data points which we consider very few.

One threat that need to be addressed in future studies is that we did not look at stable cross-functional teams in relation to social identity and how that might be different compared to the clear distinction between development groups and support groups in our sample.

\section{Conclusion and Future Work}
This paper set out to investigate if there is a connection between social identity of software engineers and software development effectiveness. The qualitative result shows that aspects of social identity affect software developers' behavior, and that we need to build cross-functional stable teams over time also from a pure social identity perspective in addition to the product related aspects to obtain high effectiveness. We believe the findings that the software developers identified strongly only with other software developers and distanced themselves from other types of roles and teams, should have an effect on their ability to deliver value to their customers, but this remains to be explored. What we do see is that DevOps and cross-functional teams make sense also from a pure social identity perspective. While we have specifically focused on in-depth analysis of interviews, the suggested quantitative approach to research social identity is of value to the software engineering community. 

In terms of future research, we particularly suggest the use of other approaches than using how many features that were delivered to a software system as a measure of effectiveness. What should really be operationalized is the added customer value even if this is difficult to achieve.

\bibliographystyle{IEEEtran}

\bibliography{refssocial}

\begin{thebibliography}{10}
\providecommand{\url}[1]{#1}
\csname url@samestyle\endcsname
\providecommand{\newblock}{\relax}
\providecommand{\bibinfo}[2]{#2}
\providecommand{\BIBentrySTDinterwordspacing}{\spaceskip=0pt\relax}
\providecommand{\BIBentryALTinterwordstretchfactor}{4}
\providecommand{\BIBentryALTinterwordspacing}{\spaceskip=\fontdimen2\font plus
\BIBentryALTinterwordstretchfactor\fontdimen3\font minus
  \fontdimen4\font\relax}
\providecommand{\BIBforeignlanguage}[2]{{%
\expandafter\ifx\csname l@#1\endcsname\relax
\typeout{** WARNING: IEEEtran.bst: No hyphenation pattern has been}%
\typeout{** loaded for the language `#1'. Using the pattern for}%
\typeout{** the default language instead.}%
\else
\language=\csname l@#1\endcsname
\fi
#2}}
\providecommand{\BIBdecl}{\relax}
\BIBdecl

\bibitem{positive_workplace}
J.~F. Defranco and P.~A. Laplante, ``Positive workplace: Enhancing individual
  and team productivity,'' \emph{PM World Journal}, vol.~3, no.~9, pp. 1--22,
  2006.

\bibitem{agile_manifesto}
M.~Fowler and J.~Highsmith, ``{The Agile Manifesto},'' In Software Development,
  Issue on Agile Methodologies, last accessed on December 29th, 2006, Aug.
  2001.

\bibitem{group_development_gren}
L.~Gren, R.~Torkar, and R.~Feldt, ``Group development and group maturity when
  building agile teams: {A} qualitative and quantitative investigation at eight
  large companies,'' \emph{The Journal of Systems and Software}, vol. 124, pp.
  104—--119, 2017.

\bibitem{group_development_wheelan_time}
S.~Wheelan, ``Group size, group development, and group productivity,''
  \emph{Small Group Research}, vol.~40, no.~2, pp. 247--262, 2009.

\bibitem{psychological_group_processes}
L.~Gren, ``Psychological group processes when building agile software
  development teams,'' Ph.D. dissertation, The University of Gothenburg, 2017.

\bibitem{social_identity_theory}
D.~Abrams and M.~A. Hogg, \emph{Social identity theory: {C}onstructive and
  critical advances}.\hskip 1em plus 0.5em minus 0.4em\relax London: Harvester
  Wheatsheaf, 1990.

\bibitem{collective_self_esteem_scale}
R.~Luhtanen and J.~Crocker, ``A collective self-esteem scale: {S}elf-evaluation
  of one's social identity,'' \emph{Personality and social psychology
  bulletin}, vol.~18, no.~3, pp. 302--318, 1992.

\bibitem{effectiveness_def}
Effectiveness., ``Cambridge english dictionary,''
  https://dictionary.cambridge.org/dictionary/english/effectiveness, n.d.

\bibitem{personal_identity_theory}
J.~E. Stets and M.~J. Carter, ``The moral self: Applying identity theory,''
  \emph{Social Psychology Quarterly}, vol.~74, no.~2, pp. 192--215, 2011.

\bibitem{identity_theory}
J.~E. Stets and P.~J. Burke, ``Identity theory and social identity theory,''
  \emph{Social psychology quarterly}, pp. 224--237, 2000.

\bibitem{identity_salience}
S.~Stryker and R.~T. Serpe, ``Identity salience and psychological centrality:
  Equivalent, overlapping, or complementary concepts?'' \emph{Social psychology
  quarterly}, pp. 16--35, 1994.

\bibitem{social_identity_perspective}
M.~A. Hogg, D.~Abrams, S.~Otten, and S.~Hinkle, ``The social identity
  perspective: Intergroup relations, self-conception, and small groups,''
  \emph{Small group research}, vol.~35, no.~3, pp. 246--276, 2004.

\bibitem{hogg2014sp}
M.~A. Hogg and G.~M. Vaughan, \emph{Social Psychology}, 7th~ed.\hskip 1em plus
  0.5em minus 0.4em\relax Harlow, England: Pearson, 2014.

\bibitem{social_formation}
T.~Postmes, R.~Spears, A.~T. Lee, and R.~J. Novak, ``Individuality and social
  influence in groups: Inductive and deductive routes to group identity.''
  \emph{Journal of personality and social psychology}, vol.~89, no.~5, pp.
  747--763, 2005.

\bibitem{khaled}
K.~W. Al-Sabbagh and L.~Gren, ``The connections between group maturity,
  software development velocity, and planning effectiveness,'' \emph{Journal of
  Software: Evolution and Process}, pp. e1896--n/a, In Press.

\bibitem{interview_preparation_conduction}
O.~Doody and M.~Noonan, ``Preparing and conducting interviews to collect
  data,'' \emph{Nurse Researcher}, vol.~20, no.~6, pp. 28--32, 2013.

\bibitem{effort_estimation}
M.~Usman, E.~Mendes, F.~Weidt, and R.~Britto, ``Effort estimation in agile
  software development: a systematic literature review,'' in \emph{Proceedings
  of the 10th International Conference on Predictive Models in Software
  Engineering}.\hskip 1em plus 0.5em minus 0.4em\relax ACM, 2014, pp. 82--91.

\bibitem{approaches_qualitative_content_analysis}
H.-F. Hsieh and S.~E. Shannon, ``Three approaches to qualitative content
  analysis,'' \emph{Qualitative health research}, vol.~15, no.~9, pp.
  1277--1288, 2005.

\bibitem{McElreath2016sra}
R.~McElreath, \emph{Statistical rethinking: {A} {B}ayesian course with examples
  in {R} and {S}tan}.\hskip 1em plus 0.5em minus 0.4em\relax Boca Raton: CRC
  Press Taylor {\&} Francis Group, 2016.

\bibitem{R}
\BIBentryALTinterwordspacing
{R Core Team}, \emph{R: {A} Language and Environment for Statistical
  Computing}, R Foundation for Statistical Computing, Vienna, Austria, 2018.
  [Online]. Available: \url{https://www.R-project.org/}
\BIBentrySTDinterwordspacing

\bibitem{carpenter2017stan}
B.~Carpenter, A.~Gelman, M.~D. Hoffman, D.~Lee, B.~Goodrich, M.~Betancourt,
  M.~Brubaker, J.~Guo, P.~Li, and A.~Riddell, ``Stan: {A} probabilistic
  programming language,'' \emph{Journal of statistical software}, vol.~76,
  no.~1, 2017.

\bibitem{van2014gentle}
R.~Van~de Schoot, D.~Kaplan, J.~Denissen, J.~B. Asendorpf, F.~J. Neyer, and
  M.~A. van Aken, ``A gentle introduction to {B}ayesian analysis:
  {A}pplications to developmental research,'' \emph{Child development},
  vol.~85, no.~3, pp. 842--860, 2014.

\bibitem{bernardo1975non}
J.-M. Bernardo, ``Non-informative prior distributions: {A} subjectivist
  approach,'' \emph{Bulletin of the International Statistical Institute},
  vol.~46, pp. 94--97, 1975.

\bibitem{gelman2008weakly}
A.~Gelman, A.~Jakulin, M.~G. Pittau, Y.-S. Su \emph{et~al.}, ``A weakly
  informative default prior distribution for logistic and other regression
  models,'' \emph{The Annals of Applied Statistics}, vol.~2, no.~4, pp.
  1360--1383, 2008.

\bibitem{2018arXiv180909849T}
R.~{Torkar}, R.~{Feldt}, and C.~A. {Furia}, ``{Arguing Practical Significance
  in Software Engineering Using Bayesian Data Analysis},'' \emph{ArXiv
  e-prints}, Sep. 2018.

\bibitem{dovidio1998intergroup}
J.~F. Dovidio, S.~L. Gaertner, and A.~Validzic, ``Intergroup bias: {S}tatus,
  differentiation, and a common in-group identity.'' \emph{Journal of
  personality and social psychology}, vol.~75, no.~1, p. 109, 1998.

\end{thebibliography}

\end{document}